\documentclass{aa}  

\usepackage{graphicx}
\usepackage{txfonts}
\usepackage[T1]{fontenc}
\usepackage{caption}
\usepackage{subcaption}
\usepackage{textcomp}
\usepackage{natbib,twoopt}
\usepackage{xcolor}
\usepackage{ulem}

\definecolor{darkgreen}{RGB}{0,100,0}
\newcommand{\newchange}[1]{\textcolor{black}{#1}}
\usepackage[breaklinks=true]{hyperref} 
\bibpunct{(}{)}{;}{a}{}{,}             
\makeatletter
  \newcommandtwoopt{\citeads}[3][][]{\href{http://adsabs.harvard.edu/abs/#3}%
    {\def\hyper@linkstart##1##2{}%
     \let\hyper@linkend\@empty\citealp[#1][#2]{#3}}}
  \newcommandtwoopt{\citepads}[3][][]{\href{http://adsabs.harvard.edu/abs/#3}%
    {\def\hyper@linkstart##1##2{}%
     \let\hyper@linkend\@empty\citep[#1][#2]{#3}}}
  \newcommandtwoopt{\citetads}[3][][]{\href{http://adsabs.harvard.edu/abs/#3}%
    {\def\hyper@linkstart##1##2{}%
     \let\hyper@linkend\@empty\citet[#1][#2]{#3}}}
  \newcommandtwoopt{\citeyearads}[3][][]%
    {\href{http://adsabs.harvard.edu/abs/#3}
    {\def\hyper@linkstart##1##2{}%
     \let\hyper@linkend\@empty\citeyear[#1][#2]{#3}}}
\makeatother
\begin{document}

   \title{Machine learning for exoplanet detection in high-contrast spectroscopy} 
   \subtitle{Combining cross-correlation maps and deep learning on medium-resolution integral-field spectra}

   \author{R.~Nath-Ranga\inst{1},
          O.~Absil\inst{1},
          V.~Christiaens\inst{1,2},
          \and
          E.~O.~Garvin\inst{3},
          }

   \institute{STAR Institute, University of Liège, 19C Allée du Six Août, 4000 Liège, Belgium\\
              \email{rakesh.nath@uliege.be}
              \and
              Institute of Astronomy, KU Leuven, Celestijnenlaan 200D, Leuven, Belgium
              \and
             Institute for Particle physics and Astrophysics, ETH Z{\"u}rich, Wolfgang Pauli Strasse 27, 8093 Z{\"u}rich, Switzerland
             }

  \date{Received 02 January 2024; Accepted 23 April 2024}

  \abstract
{The advent of high-contrast imaging instruments combined with medium-resolution spectrographs allows spectral and temporal dimensions to be combined with spatial dimensions to detect and potentially characterize exoplanets with higher sensitivity.}
  {We developed a new method to effectively leverage the spectral and spatial dimensions in integral-field spectroscopy (IFS) datasets using a supervised deep-learning algorithm to improve the detection sensitivity to high-contrast exoplanets.}
  {We began by applying a data transform whereby the four-dimensional (two spatial dimensions, one spectral dimension, and one temporal dimension) IFS datasets are replaced by four-dimensional cross-correlation coefficient tensors obtained by cross-correlating our data with \newchange{young gas giant} spectral template spectra.
  Thus, the spectral dimension is replaced by a radial velocity dimension and the rest of the dimensions are retained `as is'.
  This transformed data is then used to train machine learning (ML) algorithms.
  We trained a 2D convolutional neural network 
  with temporally averaged spectral cubes as input, and a convolutional long short-term memory memory network that uses the temporal data as well.
   We compared these two models with a purely statistical (non-ML) exoplanet detection algorithm, which we developed specifically for four-dimensional datasets, based on the concept of the standardized trajectory intensity mean (STIM) map. 
  We tested our algorithms on simulated \newchange{young gas giant}s inserted into a SINFONI dataset that contains no known exoplanet, and explored the sensitivity of algorithms to detect these exoplanets at contrasts ranging from $10^{-3}$ to $10^{-4}$ for different radial separations.
  }
  {We quantify the relative sensitivity of the algorithms by using modified receiver operating characteristic curves (mROCs).
  We discovered that the ML algorithms produce fewer false positives and have a higher true positive rate than the STIM-based algorithm.
  We also show that the true positive rate of ML algorithms is less impacted by changing radial separation than the STIM-based algorithm.
  Finally, we show that preserving the velocity dimension of the cross-correlation coefficients in the training and inference plays an important role in ML algorithms being more sensitive to the simulated \newchange{young gas giant}s.}
  {\textcolor{black}{In} this paper we demonstrate that ML techniques have the potential to improve the detection limits and reduce false positives for directly imaged planets in IFS datasets, after transforming the spectral dimension into a radial velocity dimension through a cross-correlation operation and that the presence of the temporal dimension does not lead to increased sensitivity.
  }

\titlerunning{Combining cross-correlation maps and deep learning on medium-resolution integral-field spectra}
\authorrunning{Nath-Ranga et al.}
\maketitle
%

\section{Introduction}

\newchange{Direct imaging combined with low-resolution integral field spectrographs (IFS), such as in VLT/SPHERE \citep{2019Beuzit}, Gemini/GPI \citep{2014MacintoshGPI}, and Subaru/CHARIS \citep{Groff2015}, or with medium-resolution IFS, such as in VLT/SINFONI \citep{2004SINFONI}, VLT/ERIS \citep{2023Davies}, and Keck/OSIRIS \citep{2000OSIRIS}, allows exoplanet hunters to access both images and infrared spectra of  substellar companions around nearby stars.
Since the advent of such IFS instruments, their combination with high-contrast imaging techniques has strongly contributed to the direct detection of young gas giants such as PDS\,70b \citep{2018KepplerPDS70}, or 51\,Eridani\,b \citep{2015MacintoshEridani}.
In addition, the spectral dimension provided by the IFS has allowed us to characterize directly imaged exoplanets such as HR\,8799bcde \citep[e.g.][]{2013Konopacky,Bonnefoy2016}, $\beta$\,Pic\,b \citep[e.g.][]{Bonnefoy2014,2017ChilcoteBetapic}, HD\,206893b \citep[e.g.][]{2017DelormeHD206893b}, PDS\,70b \citep[e.g.][]{2018MullerPDS70,2019ChristiaensPDS70}, or HIP\,99770b \citep[][]{2023Currie}, and has provided clues on the nature of various  low-mass substellar companions such as HD142527B \citep[e.g.][]{2018A&ChristiaensHD142527} or the as-yet-unidentified type of companion, such as CrA\,9b \citep[][]{2021ChristiaensCrA-9b}.
\newchange{The early science operations of JWST/NIRSpec \citep{2022BokerJWST}, including specific high-contrast imaging applications \citep{2023RuffioJWST}, further demonstrates the detection ability of such instruments.}}

However, these discoveries have been and far between because of the specific challenges posed by direct imaging data.
One of the more difficult challenges is to be able to separate the exoplanetary signal from the ever-present stellar glare.
The flux ratio between the exoplanet and its host star, termed contrast, can be as high as $10^{-10}$ for a Jupiter-like planet in a Solar System analogue in visible light \citep[e.g.][]{2023Galicher}.
Therefore, we work in the regime of high-contrast imaging (HCI), where the planetary flux-to-stellar flux ratio defines the sensitivity of our detections.
\newchange{In order to work in the high-contrast regime, a number of observing strategies have been employed, the most common for exoplanet imaging being angular differential imaging \citep[ADI;][]{Liu2004,2006MaroisADI} and spectral differential imaging \citep[SDI;][]{2002SparksSDI}. These two techniques can even be combined in a hybrid mode referred to as spectral angular differential imaging \citep[SADI; e.g.][]{2019ChristiaensPDS70}.}
While these strategies have their own unique features, in general, they produce a model for the stellar halo that can be subtracted from the data, leaving the residual data largely free of stellar contamination.
However, unsubtracted speckle signals in the data can still be interpreted as exoplanet companions.
Therefore, astronomers are limited to contrasts where it is possible to confidently differentiate between the exoplanet companions and speckles. These achievable sensitivity limits are generally much shallower than what would theoretically be allowed by the fundamental photon noise limit.

A possible way out of this confusion is to combine high-contrast imaging data with high-resolution spectra to improve the detectability of exoplanets, as suggested by \citet{2015Snellen}, who propose to rely on the \newchange{exoplanetary} molecular absorption lines present in the spectra.
The molecular absorption lines produced in the atmosphere of a Jupiter analogue have long been used to characterize the atmospheres of spatially unresolved exoplanets using cross-correlations with template spectra \cite[e.g.][]{snellen2010orbital,birkby2013detection} that contain molecular absorption features \citep[e.g.][]{2003BTsettl}.
The additional presence of spatial information in spectroscopic data allows astronomers to converge on the region in \textcolor{black}{the} spatial dimension where the molecular cross-correlation is the highest.
Such maps are generally referred to as molecular maps, as introduced by \citet{2018AHoeijmakersMM}, who produced detection maps for \rm{H$_2$O} and \rm{CO} in \textcolor{black}{the} SINFONI datasets of $\beta$~Pic.
In order to produce such maps, known molecular templates are cross-correlated with the data.
This cross-correlation produces coefficients that can be placed in the same pixel positions in lieu of the original photon values to produce molecular cross-correlation coefficient maps.
These molecular maps can now provide spatial and spectral characterization of the exoplanet.
\newchange{However, deriving quantitative constraints on molecular abundances in an exoplanet atmosphere from cross-correlation measurements requires using a non-trivial atmospheric retrieval framework \citep{Brogi2019}. Instead, exoplanet characterization based on medium-resolution IFS spectra has therefore generally relied on the extraction of the (possibly continuum-subtracted) exoplanet spectrum, followed by atmospheric retrieval, as performed for instance on HR$~8799$e \citep[][]{2020MolliereHR8799e}, HR$~7672$AB \citep[][]{2022WangHR7672AB}, YSES$~1$b \citep[][]{2022Zhang}, or HR$~4747$B \citep[][]{2022Xuan4747B}.}

In addition to  the challenge of retrieving abundances from cross-correlation measurements, another crucial question is how can astronomers claim a detection based on the value of the cross-correlation coefficients alone\textcolor{black}{?}
A common method of claiming a detection is to estimate whether the cross-correlation coefficient with the companion spectrum is $5\sigma$ higher than the cross-correlation noise level.
In spatial photon-based maps, such detections are usually subjected to measures such as spatial S/N \citep[][]{2014MawetSNR}.
An equivalent of such an S/N computed from the cross-correlation coefficient vector itself has been attempted by several authors \citep[e.g.][]{2018AHoeijmakersMM}, but this still suffers from an auto-correlation bias in the cross-correlation values and from the non-Gaussianity of the noise.
Some remedies for the  auto-correlation bias have been suggested by \citet[][]{ruffio2019radial}, who measure the noise separately and then divide the auto-correlation signal out.
However, these techniques also continue to rely on the assumption that noise is Gaussian in distribution.
There is also the intrinsic difficulty in measuring the S/N within a correlation pattern, as described in \cite{2023Malin} and the solutions therein. There is thus a lack of consensus in defining the $5\sigma$ limit of detection when using cross-correlation maps.
In addition, the S/N computation method used by some authors \citep[e.g. in][]{2021Cugno} requires computing large radial velocity cross-correlations, which are computationally more intensive and need to be computed one pixel at a time, which also makes them unsuited for large datasets.

\newchange{In order to tackle some of these issues, we propose the use of supervised machine learning (ML) techniques to exploit the multi-dimensional character of IFS datasets. In the literature, ML techniques have already been proposed to process the spectral and spatial dimensions separately for direct exoplanet characterization tasks. In particular, \citet{2020Fisher} demonstrated that using high-resolution spectra in an ML framework is not efficient, and proposed to feed the ML models with cross-correlation coefficients instead. 
We therefore only consider cross-correlation maps, which preserve the multi-dimensionality of the datasets, where the wavelength dimension is replaced by the radial velocity of the cross-correlation and the photon values by the cross-correlation coefficients.
The presence of these multiple dimensions \textcolor{black}{allows us to explore} the potential of using ML algorithms to claim detections as opposed to using metrics such as S/N.}
\newchange{Using spatial data with ML algorithms has had particular success in the direct imaging community, to detect point-like sources \citep{2018Gomez,2023Carlito} or to build the reference PSF to be subtracted in ADI images \citep{2022Gebhard,2023Flasseur}.}
\newchange{However, research in the field of exoplanet detection using spectral and direct imaging data with ML algorithms still has plenty of scope to develop. This is a gap that this paper, and a companion paper \citet{2024Garvin}, will be addressing.
\textcolor{black}{While in this paper} we focus on how spatial, spectral (i.e\textcolor{black}{., replaced by the cross-correlation vector}), and temporal dimensions can be most effectively leveraged together in a single ML framework. In the companion paper, an alternative approach is explored \textcolor{black}{by Garvin et al. (in prep.)} where the ML framework focuses only on the cross-correlated spectral dimension. 
This \textcolor{black}{companion paper} provides a more instrument-agnostic approach in as much as the spatial dimension is treated separately, which allows \textcolor{black}{us to use} this approach on long-slit spectroscopy datasets as well, for instance. 
\textcolor{black}{Both of these studies} aim to illustrate the as-yet-untapped potential of ML techniques to tackle spectrally resolved exoplanet datasets for exoplanet detection.
Having two parallel and complementary studies allows us to explore the potential of ML techniques in a more comprehensive manner without having to force either study to be a one-size-fits-all solution.}

\newchange{The main research question of the present paper is whether ML can improve upon the detection capability of non-ML algorithms, taking advantage of all the available dimensions in high-contrast IFS datasets. This involves addressing some intermediate questions, such as the definition of an appropriate non-ML algorithm to compare with, and the definition of performance metrics.}
To this end, this paper is organized as follows\textcolor{black}{, s}ection~\ref{sec:data} describes the data, its spatial and spectral pre-processing, the generation of cross-correlation cubes, and the insertion of simulated \newchange{young gas giants}.
Section~\ref{sec:mapbased} describes cross-correlation map-based detection algorithms, including one non-ML algorithm and two ML algorithms that we   developed for this project.
Section~\ref{sec: training ML} describes the important process of preparing our data to train the ML algorithms.
Section~\ref{sec:results} describes how results from the ML and non-ML algorithms can be evaluated with the same baseline comparison methods.
Finally, in Sect.~\ref{sec:discussion} we discuss the reasons why using data in specific ways allows us to achieve higher detection sensitivity, and conclude with the limitations of our study.\footnote{The code used in this paper is available at \url{https://github.com/digirak/NathRanga2024.git}.}\\
\section{From 4D IFS tensors to cross-correlation tensors}\label{sec:data}

This section describes the step-wise transformation of the data from 4D IFS (i.e. 2D spatial, 1D spectral and 1D temporal) tensors to cross-correlation tensors that can then be used to train ML algorithms.
We start with describing these 4D tensors in Sect.~\ref{sec:datadesc}.
We follow by describing the injection of simulated \newchange{young gas giant}s in this data in Sect.~\ref{sec: FC insertion}.
Finally, we describe the conversion of these 4D IFS tensors into 4D cross-correlation tensors in Sect.~\ref{sec:specpreproc}.

\subsection{Observations and data calibration}\label{sec:datadesc}

The ESO pipeline calibration includes flat-fielding, wavelength calibration, detector linearity correction and extraction of spectral cubes from each raw detector frame. 
For this study, we used a dataset obtained on HD~$179218$ with the SINFONI instrument on the VLT \citep{2004SINFONI,2003SEisenhauer} as part of ESO program 093.C-0526.
We chose this dataset because of the absence of any known companion.
The data was calibrated as in \citet{2018A&ChristiaensHD142527}, using the SINFONI pipeline implemented in the ESO common pipeline library \citep[EsoRex version 3.10.2;][]{2006Abuter}, apart from the sky subtraction, which was performed manually before running the ESO pipeline.
The ESO pipeline calibration includes flat-fielding, wavelength calibration, detector linearity correction and extraction of spectral cubes from each raw detector frame. 
Each spectral cube is composed of $\sim 2000$ channels spanning from $1.45\mu$m to $2.45 \mu$m (i.e. $H$ and $K$ bands).
This is the spectral dimension.
Each of the calibrated spectral cubes are then used as input for further preprocessing.
Each spectral frame is made of $64\times64$ pixels with a plate scale of $12.5$ mas/pixel. However, due to a twice coarser sampling vertically, each pair of consecutive rows have the same values.
This forms the spatial dimension.
We have a sequence of $83$ such spectral cubes captured in pupil-tracking mode, with the instrument rotator turned off.
This constitutes the temporal dimension.
Each of the calibrated spectral cubes are then used as input for further preprocessing.
\newchange{The wavelength vector is recalculated by cross-correlating the sky signal present in spectral channels at the edges of the H and K band channels with the expected ESOCalc transmission profile following \cite{2018AHoeijmakersMM}. }
All of the pre-processing is carried out in accordance with the recipe set out in the SINFONI data reduction pipeline \citep{2006Abuter}.
We performed basic pre-processing of the data in order to:
\begin{itemize}
    \item correct bad pixels in all images,
    \item determine the centroid location of the star and
    \item produce the final wavelength solution of the cube. 
\end{itemize}
We identify bad pixels with an iterative sigma-clipping algorithm, and correct them with a 2D Gaussian kernel, using routines of the Vortex Image Processing (\textsc{VIP}) package \citep{2017AJGomezVIP,2023Christiaens}.
We resample each image vertically by a factor two, to avoid the redundant rows. 
We find the stellar centroid by fitting a 2D Gaussian model to each spectral channel. 
These fits also enable us to estimate the FWHM of the PSF in each channel. 
In this paper we explore the detection of a population of planets which are relatively young (typically a few tens of Myr old) where they still retain a large fraction of their formation entropy, with a $T_{\rm eff}\approx 1000 - 1500$~K.
These exoplanets are called \newchange{young gas giant}s because of their \newchange{age} and gas giant composition.
In order to simulate the presence of a \newchange{young gas giant} in the data we inject a simulated \newchange{young gas giant} at a specific radial distance and position angle (PA) from the frame centre.
We use the substellar templates characterized by their effective temperature ($T_{\rm{eff}}$) and surface gravity ($\log(g)$), and contain detailed spectra which model the atmosphere for brown dwarfs, late M-type stars, and young gas-giant planets or \newchange{young gas giant}s.
\newchange{Young gas giant}s are particularly interesting as they seem to be present in the $5-20$~au distance range \cite{2016Bryan}.
For this paper we use the models from BT-SETTL as \newchange{it allows us} to have a large enough wavelength range in the infrared. 
With the BT-SETTL models we have access to wavelengths between $1\mu$m to $30\mu$m.
We choose a \newchange{young gas giant} template with a $T_{\rm eff} = 1300$K and a $\log(g)=3.0$.
The wavelength vector is recalculated by cross-correlating the sky signal present in spectral channels at the edges of the H and K band channels with the expected ESOCalc transmission profile following \cite{2018AHoeijmakersMM}. 
\newchange{The spectral frames are then cropped around the identified stellar centroid positions,which falls typically at $\approx 2-3$ pixels from the centre of the frames during the observation. This step both recentres the images on the star and removes artefacts that affect spaxels at the edge of the frames.
This in turn reduces the field of view of the frames to $61\times61$ pixels.}
Finally, we collate all re-centred cubes into a single master 4D tensor.
The tensor now has $83$ temporal cubes with $2000$ wavelength bins with $61$ rows and $61$ columns of pixels.

\subsection{Injection of simulated \newchange{young gas giant}s in the data}\label{sec: FC insertion}

In order to simulate the presence of a \newchange{young gas giant} in the data, we inject a simulated planet using a template spectrum at a specific radial distance and position angle (PA) from the frame centre.
In this paper we use the template set provided by the BT-SETTL models \citep{1997Allard, 2011Allard,2003BTsettl}.
We use substellar templates characterized by their effective temperature ($T_{\rm{eff}}$) and surface gravity ($\log(g)$), containing detailed model spectra for the atmosphere of brown dwarfs, late M-type stars, and young gas-giant planets.
For this paper we use the BT-SETTL models \newchange{they allows us} to have a large enough wavelength range in the infrared, ranging between $1\mu$m and $30\mu$m.
For most of our test, we choose a \newchange{young gas giant} template with a $T_{\rm eff} = 1300$K and a $\log(g)=3.0$.
When injecting this template, we follow these steps:
\begin{enumerate}
    \item the template is re-sampled to the same spectral resolution as SINFONI and the wavelength range is limited to that of SINFONI (i.e. between $1.4$ to $2.4$ $\mu$m);
    \item we broaden each of the bins in the template spectrum to match the spectral PSF width of the SINFONI instrument (derived from the user manual);
    \item we remove the spectral bins between $1.8$ $\mu$m and $1.91$ $\mu$m corresponding to the telluric absorption range, and then re-normalize the spectrum so that the total number of photons in each spectrum is $1$;
    \item we then multiply the total photon count in the spectrum by a factor of the total photon count in the spectrum extracted at the located of the star in our IFS dataset (this factor is the contrast);
    \item finally, we chose a target pixel and add a PSF-scaled equivalent of the \newchange{young gas giant} spectrum in a FWHM-sized aperture around that pixel.
\end{enumerate}
We use routines from the \textsc{VIP} package to perform the spatial injection at a specified radial separation and position angle, and at a given contrast with respect to the star.
Every insertion in each temporal cube is done taking into account the parallactic rotation of the frame (or spectral cube in this case).
In order to rule out the possibility of a real \newchange{young gas giant} biasing our tests, the simulated planets are inserted using the opposite parallactic angles to the ones associated with the data, and these are later used for final derotation of the frames.

\subsection{Producing cross-correlation cubes}\label{sec:specpreproc}

After injecting fake planets into our 4D IFS data cubes, we now proceed with the computation of the cross-correlation, which consists of two steps: (i) reducing the amount of stellar feature contamination in the data, and (ii) cross-correlating each spaxel with a similarly pre-processed version of a BT-SETTL planetary template with similar $T_{\rm eff}$ and $\log(g)$ to the one used for the injection.
The idea of using a similar but not exact same template is to take into account the possible mismatch between BT-SETTL models and real planets, and avoid a too optimistic set of cross-correlation tensors. 
\newchange{The chosen templates are typically picked two grid points away from the injected template. This is expected to preserve a relatively high cross-correlation with the injected template, as cross-correlations are not very sensitive to $T_{\rm eff}$ and $\log(g)$. 
Picking templates further away from the injected template within the BT-SETTL grid may result in unrealistically low cross-correlation values. In practice, when performing planet detection in a yet-undetected planetary system, a large number of templates will generally be used, which is expected to result in relatively high cross-correlation in at least a few cross-correlation tensors. This is the situation that we try to reproduce here.}

\subsubsection{Stellar spectral feature subtraction}

To remove any lower order residual contamination, we compute \newchange{the spectrum filtered using the Savitzky-Golay filter \citep{1964SavitzkyGolay}} with a window size of $101$ and subtract it from the extracted spectrum. The original spectrum is then replaced by this residual spectrum at the same pixel.
Before computing the cross-correlation on each spaxel of our IFS cubes, it is important to first reduce the amount of stellar light in the data. 
While classical HCI post-processing techniques such as ADI or SDI rely on the spatial structure and behaviour of stellar speckles to remove them from the data, here we rely on the spectral behaviour following the first step of the High Resolution Spectral Differential Imaging (HRSDI) introduced by \citet{2019Haffert}. 
This step consists of constructing a reference spectrum that contains the stellar spectral features and then dividing \citep[or subtracting, in the case of][]{2019Haffert} this reference spectrum from spectra extracted from every pixel using aperture photometry. 
To produce a reference spectrum we have the following sequential steps:
\begin{enumerate}
    \item we first extract the stellar spectrum using aperture photometry at each frame centre with an aperture radius of $1$ FWHM for each wavelength for every temporal cube;
    \item we then normalize this extracted spectrum by the total photon counts for every wavelength in every cube;
    \item and finally we remove the telluric bins ($1.81$ $\mu$m to $1.93$ $\mu$m) by masking the values in these bins.
\end{enumerate}
This spectrum is the reference spectrum, which we compute for every temporal cube so that we end up with as many reference spectra as temporal cubes (in our case $83$ spectra).
We then divide the rest of the spectra extracted from each of the spaxels in each temporal cube by its own reference spectrum.

\subsubsection{Cross-correlation of the template with target spectrum}\label{sec: CC algorithm}

Cross-correlation is a method used to compare two different spectra.
In this paper, such a comparison is made between an extracted spectrum and the BT-SETTL template spectrum, which is downsampled and broadened to match the SINFONI spectra (same as when it was injected) and then mean-subtracted as described above.
Cross-correlation is the result of introducing a small velocity shift between the model and the extracted spectrum and computing the product of the two.
This product is the cross-correlation coefficient and it is computed for every velocity shift.
The cross-correlation $\mathcal{C}$ between the model template $M_{\lambda}$ and the extracted spectrum ($F_{\lambda}$) at different velocities $v$ is thus given by
\begin{equation}
    \mathcal{C}(v) = \sum\limits_{\lambda}F_{\rm{\lambda}}\times (M_{v,\lambda}-M_{\rm{SG},v,\lambda})  \; ,
    \label{eq:CC equation}
\end{equation}
where $v$ is the velocity shift, $M_{v,\lambda}$ the velocity-shifted version of $M_{\lambda}$, and $M_{\rm{SG},\lambda}$ the model template after applying the Savitzky-Golay filter.
In this paper, we sample the velocities at every $10$ km/s between $-100$ to $100$ km/s, resulting in a total of 20 velocity bins. Thus, for each pixel, $F_\lambda$ is now replaced by $\mathcal{C}(v)$. This lowers the total size of the data although the number of data dimensions remain the same, and we have a 4D tensor once again with the dimensions $\left(83\times20\times61\times61\right)$ where the 20 corresponds to $v$ replacing the 2000 wavelength bins.

\section{Detection algorithms}
\label{sec:mapbased}

Cross-correlations as defined in Sect.~\ref{sec: CC algorithm} are typically used to compare spectra, by comparing the value of the cross-correlation coefficient computed at $v=0$ (i.e. $\mathcal{C}(0)$) to the standard deviation of the cross-correlation signal $\mathcal{C}(v)$ at higher velocity shifts.
This ratio is then used to define the signal-to-noise ratio (S/N) for the cross-correlation.
This, however, biases the noise estimation for a cross-correlation because of the presence of auto-correlation.
Thus, the noise is typically overestimated. 
However, if the standard deviations are computed at sufficiently high velocity shifts from the central peak \citep[e.g. for $|v|\ge 250$ km/s in][]{2018AHoeijmakersMM} then this bias is somewhat mitigated.
Even then, \citet{ruffio2019radial} show that this problem is not completely mitigated. 

Direct imaging data as used in this paper allows us to produce cross-correlation maps for a simulated \newchange{young gas giant} as described in Sect.~\ref{sec:specpreproc}.
The image dimension gives us access to neighbouring pixels where the noise of the cross-correlation can be measured as suggested for example by \citet{2022Patapis}.
All of these methods rely on computing a large set of cross-correlations for every spaxel making the problem computationally intensive.
An alternative way is to use the spatial S/N definition, that does not rely on the cross-correlation noise at all, and is commonly used in ADI algorithms as defined in \citet{2014MawetSNR}.
Instead of using this S/N definition, which once again relies on some definition of spatial noise distribution,and thus implicitly making assumptions about noise distributions,  we choose to adapt the standardized trajectory intensity mean \citep[STIM,][]{2019Pairet} map to our cross-correlation cubes. 
A major advantage of using STIM is that it makes use of the temporal axis in the residual cubes, which has shown to reduce false positives in ADI cubes.
\subsection{Standardized trajectory correlation mean (STCM) maps}

Detecting an exoplanet in an imaging dataset requires the estimation of the signal versus the noise distribution of the dataset, and tools such as S/N maps provide a means of doing this as demonstrated by \citet{2014MawetSNR}.
Here, we introduce the standardized trajectory correlation mean (STCM) map as an analogous way to estimate the relative strength of the signal. To produce an STCM map we perform the following steps.
\textcolor{black}{We first compute the cross-correlation tensor as described in Sect.~\ref{sec: CC algorithm} and derotate the cubes so that the pixels at which the simulated young gas giant was inserted now align in all the cubes.
Next we choose the velocity at which $\mathcal{C}(v)$ is the highest for this pixel\footnote{Since we do not apply an explicit frame shift and both the template and the spectrum are at the same rest frame, the highest $\mathcal{C}(v)$ should be at $v=0$ in principle.} and recompose the tensor into a cube consisting of the spatial and the temporal dimensions only.
Finally the STIM algorithm is applied to this cube along the temporal axis, resulting in an STCM map.}

\begin{figure*}
    \centering
    \includegraphics[width=0.8\textwidth]{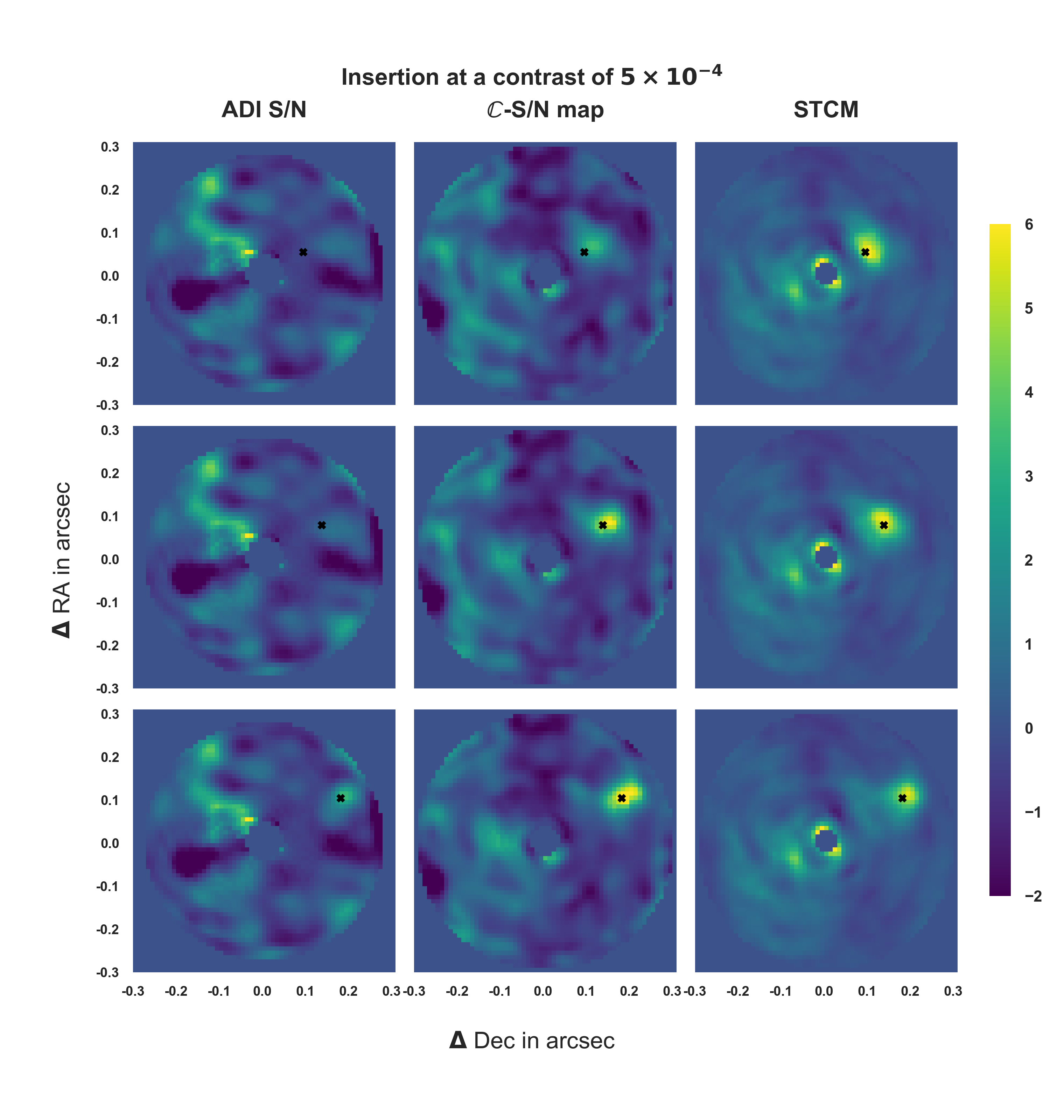}
    \caption{Detection maps obtained with different algorithms, where a \newchange{young gas giant} was inserted at a specific contrast at different separations from the frame centre. Column 1 corresponds to a classical ADI algorithm, where an intensity map is computed for each wavelength, and the S/N map is derived from the median of all the ADI-wavelength maps. Column 2 is obtained by computing an S/N map on the cross-correlation map obtained with $v=0$ km/s. Column 3 represents the STCM map, produced as described in the text. The rows represent maps produced when the same companion is inserted at different radial separations from the frame centre, respectively at $2.3$, $3.3$, and $4.3$ FWHM.
    The simulated \newchange{young gas giant} is inserted at the same contrast of $5\times10^{-4}$ in all maps.}
    \label{fig:fig_1}
\end{figure*}

In Fig.~\ref{fig:fig_1} we demonstrate the application of three different ways of estimating the relative signal strength of the simulated \newchange{young gas giant} that we insert in the data as described in Sect.~\ref{sec: FC insertion}.
Each column in Fig.~\ref{fig:fig_1} is the detection map for an injection at a fixed contrast of $5\times 10^{-4}$ produced using different algorithms, with each row corresponding to a different radial separation for the injection.
The first column is produced when we compute an S/N map \citep[as defined in][]{2014Mawet} after performing classical ADI on each 4D tensor.
In order to do this we compute the ADI for a series of data cubes at every wavelength. Then, we compute the mean ADI cube from this result. 
The second column is produced by calculating the S/N map \citep[as defined in][]{2014Mawet} on the cross-correlation cube in two steps: first we compute the median along the temporal dimension (after applying appropriate de-rotation), and then we compute the cross-correlation map and apply the standard spatial S/N to it.
The third column is produced by the STCM algorithm.
The range of values on each of these maps is normalized to be between $-2$ and $6$ so that we get a fair visual estimation of the noise values produced by each of these algorithms, based on the global minima and maxima of these nine maps.
From the figures it appears that the STCM detection map recovers the exoplanet more consistently across different separations. We interpret this as being caused by the original STIM algorithm design, which aims to alleviate small sample statistics to perform better than standard S/N maps at small separations.

\subsection{Map-based detection through deep-learning algorithms}\label{sec:ML algorithms}

Application of artificial neural networks and particularly deep neural networks to exoplanet detection is now an established method in astrophysics \citep[][]{2020Fluke}.
The exoplanet community has benefited from the use of deep neural networks for Kepler light curves \citep{2018Pearson}, for direct imaging detection \citep{2018Gomez}, and to accurately produce a simulated model of the PSF of an instrument using half-sibling regression \citep{2022Gebhard}.
We seek to use deep-learning algorithms to produce maps akin to the STCM maps described in the previous section.

The building block of a neural network is a neuron.
A neuron consists of an input that is operated on by an activation function, resulting in an output.
The activation function is usually a non-linear mathematical operation (for instance a hyperbolic tangent) that is performed on the input.
A neural network usually contains more than one neuron, and a set of such independent neurons each operating on a part of the input is called a layer. 
If the output of this layer is used as input to another layer, this layer is called a hidden layer.
When one or more hidden layers are present, such a neural network is called a deep neural network.
A layer of neurons can have multidimensional input: typically a deep neural network is used with one dimensional data and a two dimensional deep convolutional neural network is used with image data.
Typically, memory-based networks such as a long short-term memory and transformers can be used with data that have some temporal coherence.
Deep neural networks have the versatility to also use images through 2D convolutional neural networks \citep[CNN,][]{shi2015convolutional} and to combine images with temporal data in convolutional long short term memory \citep[convLSTM,][]{1997HocherieterLSTM,2022convLSTM}.
Deep neural networks have been applied to astronomical spectra \citep[e.g.][]{2019Leung,2020Tao} to classify them and specifically in direct imaging.
\citet{2020Fisher} characterized exoplanets using cross-correlations of the spectra from the HARPS-N spectrograph. A neuron, despite its input dimensionality, has a set of weights and biases as parameters. 
These parameters form the model of the neural network and are produced when we train the algorithm on data.
Training is the term used to make the neurons identify intrinsic relationships between different data dimensions and the necessary output.
In this case we train the neural networks in this paper to predict the presence of an exoplanet in the input cross-correlation tensor.
Training typically consists of providing the neural networks with large number of labelled examples of the classes one wishes to predict.
The neural network in turn predicts the probability of each class given the input data.
Training neural networks is very important to achieve the desired result; we explain this in detail in Sect.~\ref{sec: training ML}.

In this section we describe the two different deep neural network architectures that are used in this paper.
Architectures are typically defined by the number of neurons, number of layers, and the kind of neurons used.
Thus, while describing the deep-learning algorithms we have two parts, the first being its architecture, which is necessary to explain the neuron type, its activation function, and its output, and the second being its training methodology.
We explore the detectability of \newchange{young gas giant}s with different dimensionalities of the data with both a CNN-based deep neural network and a convLSTM-based network.
When using the CNN-based network, we compute a temporal mean reducing the dimensionalities to $x$, $y$, and $v$.
For the spatial-velocity dimension, we use a CNN-based deep-learning framework called Cross-Correlation and Convolutional neural network-based exoPlanet detectOr (C3PO).
In order to include the temporal dimension, we also introduce a 2D convolutional long short-term memory (LSTM)-based algorithm called Cross-correlation and convolutional LSTM based exoplANet DetectOr (C-LANDO).
We have designed these algorithms so that they are able to produce probability maps akin to the STCM maps. 
The input to these deep-learning algorithms will be the cross-correlation tensors provided pixel by pixel and the output will be the probability of that pixel containing an exoplanet. 

\subsubsection{C3PO}

\begin{figure}[!t]
    \centering
    \includegraphics[height=0.6\textheight]{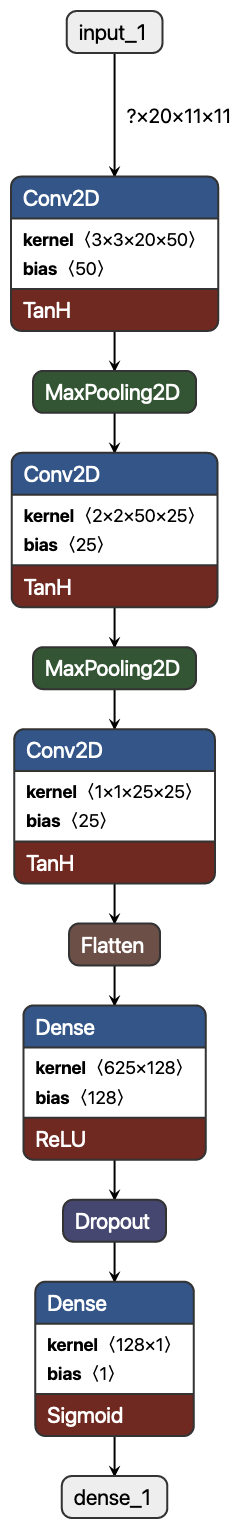} \hspace*{5mm}
    \includegraphics[height=0.55\textheight]{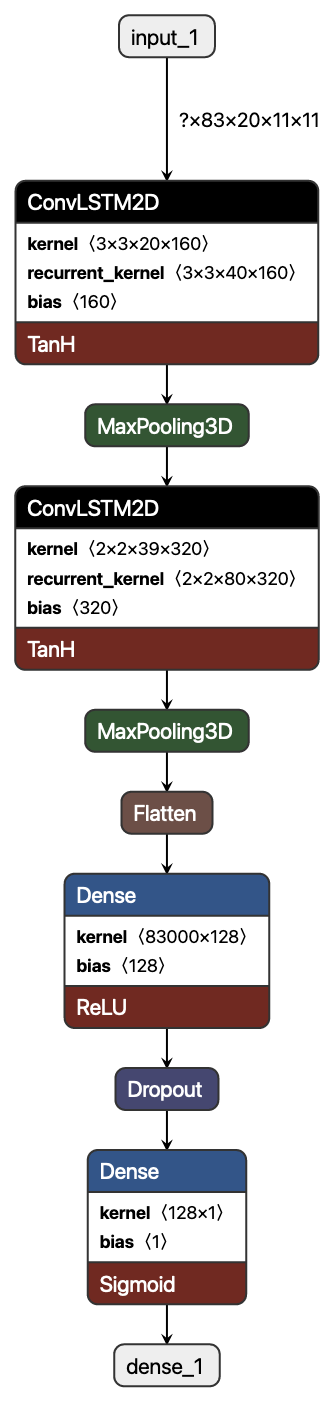}
    \caption{Schematic of the C3PO (left) and C-LANDO (right) architectures, showing the different layers and sizes of the input, the dilation of this input at different layers, and finally the output format.
    The top grey block represents the input, with dimensions printed next to the arrow (e.g., for C3PO: $20$ velocity bins, $11\times11$ pixel image size as explained in Sect.~\ref{sec: training ML}, and batch size given as input labelled with a question mark).
    Each block represents a layer and has three parts: the blue or black part represents the type of neuron, the white part is the number of neurons represented by a convolutional kernel size and number of bias units, and the red part represents the non-linearity (hyperbolic tangent).
    The kernel has four dimensions:   the first three represent the input shape and the last  represents the depth of the kernel.
    The kernel shape also represents the output dimensions of a layer and the input to its following layer. 
    Between each layer we have pooling layers marked in green, a flatte\textcolor{black}{n} layer in brown, and a dropout layer in the end. 
    The output is just a single neuron with a sigmoid activation (dense\_1).}
    \label{fig:c3po schematic}
\end{figure}

Our architecture consists of a set of convolutional layers and set of dense layers finally terminating in a single neuron, which provides a probabilistic output.
The $x$ and $y$ dimensions of the convolutional layer is called kernel size.
Several such kernels are stacked in a 3D block and the depth is the number of such kernels that are stacked.
The input image is passed through each of these layers in a block for each feedforward pass.
In our case, an input consists of the cross-correlation coefficients in the $x$, $y$, and $v$ dimensions.
The pooling layer downsamples the kernels of the convolutional layer based on a summary statistic of the output of the convolutional layer (e.g. maximum value).
The output of this layer is then flattened to remove 3D elements and then connected to a multi-layer perceptron, with a sigmoid activation as final layer. 
The latter will act as the classifier output, which is the probability of a pixel to contain a simulated \newchange{young gas giant}. The detailed architecture of C3PO is as follows:
\begin{enumerate}
    \item a convolutional layer with kernel size of $\left(3,3\right)$ with $50$ such kernels with a hyperbolic tangent activation function;
    \item a pooling layer that uses the maximum value statistic with a pool size $(2,2)$ such that  at this point the data dimensionality follows the dimensions set by C3PO, and  we now have a new dimensionality to the output based on the kernel size, number of velocity bins, and number of such kernels;
    \item a convolutional layer with kernel size $\left(2,2\right)$ with $25$ such kernels with the same activation function;
    \item another pooling layer of size $(1,1)$;
    \item a final convolutional layer of kernel size $(1,1)$ with $25$ such kernels;
    \item a fully connected layer with $128$ neurons with a rectifying linear activation function, followed by a dropout and final neuron with a sigmoid activation.
\end{enumerate}
The architecture of C3PO is illustrated in Fig.~\ref{fig:c3po schematic}. We provide more details on the input dimensionalities in Sect.~\ref{sec: training ML}.

\subsubsection{C-LANDO}

The basic unit of C-LANDO is a convolutional LSTM, in the same way as the basic unit of C3PO is a CNN. A convolutional LSTM is a combination of the standard LSTM \citep{1997HocherieterLSTM} with a convolutional neural network \citep{1990ZhangCNN}, which allows this kind of network to combine C3PO features of velocity and spatial PSF structure with the temporal structure embedded in the $83$ temporal cubes that are present. 
Our convolutional LSTM architecture intends to exploit the recurrence of spatial and velocity signatures in each temporal cubes, in the presence of a planet.
The architecture of C-LANDO, illustrated in Fig.~\ref{fig:c3po schematic}, consists of the following layers:
\begin{enumerate}
    \item a convolutional LSTM layer consisting of a $3\times3$ kernel with $40$ filters 
    \item  this is followed by a 3D max pooling layer, which pools the convolutional kernels of the LSTM two pixels at a time;
    \item this is followed by another convolutional LSTM layer consisting of a $2\times2$ kernel with $80$ filters once again followed by a max pooling layer as the previous layer, which once again pools the convolutional kernels two pixels at a time;
    \item  this is followed by a flatten layer which converts the 3D shape into 1D that can then be fed to a feedforward neural network layer  with $128$ neurons with a Rectifying Linear unit (ReLu) activation;
    \item the output of the feedforward neural network is then fed into a dropout layer that randomly drops $25\%$ of the neurons;
    \item finally, the output is a single sigmoid neuron that gives the probability of our output class.
\end{enumerate}


\section{Training the deep-learning algorithms}\label{sec: training ML}
At the beginning of training we initialize the parameters of the neural networks such as weights and biases to random values, which will then be tuned through the training process. 
An important concept that will be used repeatedly is the loss function (hereafter, referred to as ``loss''), which depends on the output type, and quantifies the difference between the ground truth and prediction of a network.
For example, in our case where the output is a binary classifier, the categorical cross-entropy loss is used.
This loss is back-propagated through the network each time the data is processed (passed) through the network \citep{Rojas1996}.
At this time the network parameters such as the weights and the biases are modified slightly and the loss is once again computed for the next data point.
In practice, a series of labelled examples (known as data samples) are input and passed through each layer of the neural network and the error between the output of our architecture and the label is computed for every example.
Typically, this process is repeated once through the whole dataset and such a training step is called an epoch.
In order to complete the training, several such training iterations or epochs have to be completed.
Each time the error is back-propagated through the network, the weights and biases are adjusted so that the error is minimized.

After calculating this loss over multiple epochs, the network parameters undergo significant adjustments to ensure that the subsequent processing by the network does not result in further loss reduction. At this point, the network is referred to as a trained model.
Training the model necessitates the generation of diverse dataset, a crucial step to prevent the network from memorizing recurring noise features and to ensure the dataset encompasses a sufficient array of data features for the network to learn from. The consideration of these challenges is pivotal in the training process, as the potential problems in the data, such as overfitting \citep[][]{dietterich1995overfitting} and gradient stall \citep[][]{patel2017sgd}, could arise if not effectively addressed.
Hence, in the following section, we  outline the process of generating training data and provide a detailed description of the training process for both C3PO and C-LANDO.
\subsection{Generating the training dataset}

\begin{figure*}[t]
    \includegraphics[width=\textwidth]{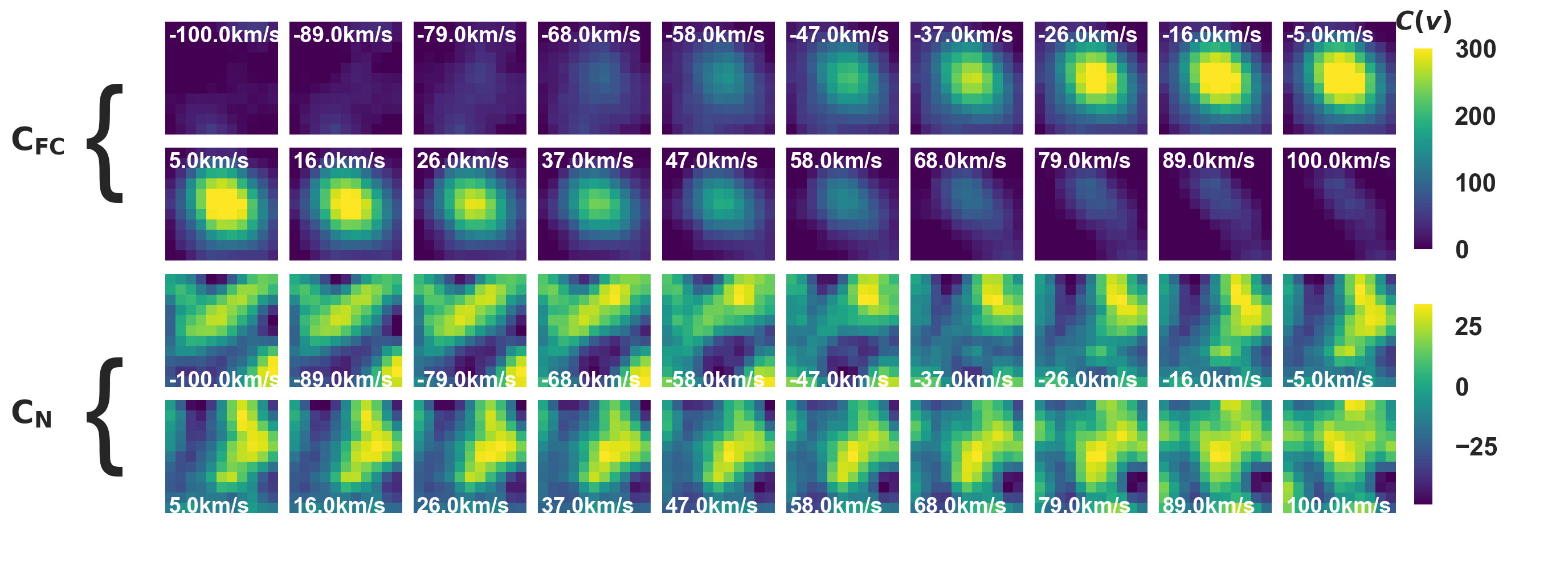}
    \caption{
    Evolution of spatial crops with respect to radial velocity. The rows indicate the different velocities and each square shows the spatial noise and signal diversity of the patch. 
    This is just a single spatio-temporal sample (i.e. one pixel from a single temporal frame).
    The colour-coded pixel value corresponds to the cross-correlation signal from Eq.~\ref{eq:CC equation}. 
    }
    \label{fig:fig-2}
\end{figure*}

To train the C3PO and C-LANDO networks presented in Sect.~\ref{sec:ML algorithms}, we insert simulated \newchange{young gas giant}s at different selected pixels with a range of contrasts and at different PAs from the frame centre, as described in Sect.~\ref{sec: FC insertion}.
In practice these contrasts range from $10^{-2}$ to $10^{-4}$ and we insert these at each pixel with a PA sampling rate of one insertion for every $30^{\circ}$ of PA, to make sure that at the closest radial separations we still have a spaced out PA separation between insertions. 
From previous studies using deep learning to detect exoplanets in images \citep[e.g.][]{2018Gomez,2023Carlito}, we learned that using full-frame images to train deep neural networks on spatial features is not optimal, as the features corresponding to an exoplanet are confined to a few percent of the total pixels that we train on. 
The effect of such imbalance can result in sub-optimal training of ML algorithms.
Therefore, we crop the field of view that is used as an input to the neural network to a region of two FWHM around the inserted fake companions (or around another position to learn the noise structure of non-detections).
This results in data with the following dimensions: \textcolor{black}{a temporal dimension of $83$ temporal cubes, a velocity dimension composed of $20$ velocity shifts starting from $-100$ and extending up to $100$ km/s and two spatial dimensions of two FWHM size each ($11\times 11$ pixels).}
Thus, each pixel now corresponds to a tensor of these dimensions as an input to C3PO and C-LANDO.
Both C3PO and C-LANDO produce as output a probability based on their input and that output is interpreted as the probability of a simulated \newchange{young gas giant} being present at central pixel of the input tensor.
We use VIP to spatially crop each full frame into non-overlapping smaller $11\times11$ pixel sized patches.

Patches are produced for a pixel where the simulated \newchange{young gas giant} is inserted, but also for those pixels where nothing is present.
The patches that contain the simulated \newchange{young gas giant} are part of the $C_{\rm{FC}}$ class (where FC stands for fake companion) and the rest, which are just noise samples, become part of the $C_{\rm{N}}$ class (where N stands for noise).
Figure~\ref{fig:fig-2} shows a sample for both $\rm{C_{FC}}$ and $\rm{C_N}$ side by side.
Spatially the $\rm{C_{FC}}$ has an intensity distribution very similar to the PSF shape originally measured in the data whereas the intensity distribution in $\rm{C_N}$ class appears to be random.
The range of cross-correlation values in the $\rm{C_{FC}}$ and the $\rm{C_N}$ also offer a point of difference.
The first two lines are those representing $\rm{C_{FC}}$ and the last two rows are produced in the same manner with $\rm{C_N}$.
In any data cube if an injection is made at a specific PA and radial separation and then the cube is cropped down into $2$ FWHM sized sub-cubes, we will have a total of $35$ non-identical $\rm{C_N}$ and $1$ $\rm{C_{FC}}$ patch tensor for every cross-correlation tensor (produced from the insertion at the same PA and radial separation) that have the spatial, velocity and temporal features. 
Diversity in the $\rm{C_{FC}}$ class is achieved by inserting the simulated \newchange{young gas giant} at different separations from the frame centre and at different contrasts to the stellar flux and then producing cross-correlation tensors.
The temporal dimension (not shown here) is such that each pixel in the $\rm{C_{FC}}$ class that contains the exoplanet shows consistently the same pixel value and a set of pixels tend to maintain a PSF-like shape for a fixed velocity. 
The features that we intend C3PO to learn are this spatial variation and the velocity variation which will be learnt as a `colour' by the CNN. 
C-LANDO will have the same features in addition to the repeating format temporally of $\rm{C_{FC}}$ and more random variation temporally.

In order to produce diverse noise samples, we resort to data augmentation similar to what was done by \citet{2018Gomez}.
Random patch sequences are subjected to one of the following augmentation techniques during insertion and before producing the $C_{N}$ patch tensors themselves: \textcolor{black}{random rotations between $0^{\circ}$ and $360^{\circ}$, random subpixel shifts of up to $0.5$ pixels in either $x$ or $y$ direction and production of additional samples by adding two non-sequential and randomly chosen samples.}
Spatially, a $C_{\rm FC}$ sample has an intensity distribution very similar to the PSF shape originally measured in the data whereas the intensity distribution in $C_{\rm N}$ class appears to be random.
The range of cross-correlation values in the $C_{\rm FC}$ and the $C_{\rm N}$ also offer a point of difference.
This is the type of differentiating feature that allows a deep-learning algorithm to make fine distinctions.
Such a differentiating feature was observed in the PCA dimension by \citet{2018Gomez}.
The temporal dimension (not shown here) is such that each pixel in the $C_{\rm FC}$ class that contains the exoplanet shows consistently the same pixel value and a set of pixels tend to maintain a PSF-like shape for a fixed velocity. 
\subsection{Training the deep-learning algorithms}

As stated earlier, ML algorithms always run the risk of memorizing the data, but there are several ways to overcome this.
One of the several strategies is to divide the data into three independent datasets called training, validation, and testing.
Overfitting is characterized by exceptionally good performance in the training dataset but conversely produces poor performance in the validation and test dataset.
The training dataset is used to primarily train the algorithm to make predictions, and the validation is used to verify the training accuracy and make small modifications to hyperparameters such as learning rate, number of neurons, and activation functions. 
The test data is reserved to test and benchmark the algorithm and is typically used when the training and validation accuracies are very close to each other.
The three datasets are typically separated before beginning training so that the algorithm never sees the other two parts when training.
The testing samples are first removed from data by choosing $25\%$ of the PA values and removing all $C_{\rm FC}$ and $C_{\rm N}$ samples corresponding to those. 
For the purposes of this paper, this corresponds to four PA values.
Then we split the remaining data into $90\%$ training and $10\%$ validation.
A key metric we use in the validation data is the number of false positives the algorithm generates. 
The validation comprises of patch tensors corresponding to just one PA value with all of the different radial separations so that we probe the different noise levels.
As with the test, the $C_{\rm N}$ and $C_{\rm FC}$ are removed before training.
We describe the computation of false positives in Sect.~\ref{sec:results}.
We use the number of false positives in the validation data to tune the mix of contrasts and the parameters of the ML algorithms, such as the  number of layers, neurons.

We started with a small network and grew it layer by layer as the validation data produced more true positives (i.e. started detecting the simulated \newchange{young gas giant} over multiple separations and for higher contrasts) and fewer false positives.
It became quickly clear that hyperparameters of the network (such as the  number of neurons, number of layers, and activation functions) themselves have little impact on the number of false positives produced in the validation data.
Initially, we started with the full range of contrasts starting from $5\times10^{-2}$ to $10^{-4}$ with no subpixel shifts.
We noticed that, as we increase the number of high-contrast examples in the data, we encounter higher false positives generated. Therefore, we limit the contrast to $<10^{-3}$ \newchange{(i.e. to simultated new gas giants that are brighter than $10^{-3}$)} with the full range of separations starting from $\sim 1$ FWHM up to $\sim 5$ FWHM.
We use the $\texttt{binary\_cross\_entropy}$ loss function \citep{1993Li} to compare the prediction of the deep-learning algorithms with the truth value.
We use the Adam optimizer \citep{2014Adam} to minimize the loss and overcome local minimas.

\section{Comparing the performance of the algorithms}\label{sec:results}

We benchmark all our algorithms based on their ability to accurately detect simulated \newchange{young gas giant}s at contrasts dimmer than $10^{-3}$ with the smallest number of \textcolor{black}{the fewest number of mistaken detections.}
\newchange{At this point, it is important to make the distinction between the contrasts supplied when training the ML algorithms, which  is limited to simulated young gas giants brighter than $10^{-3}$ to have less overfitting, while during testing we used the full range of contrasts (i.e. $10^{-2}$ to $10^{-4}$).}
In this section, we describe the production of detection map and binary maps with our test data described in the previous section.
The binary maps are then used to compute true and false positives, and we finally describe the metric used for the comparison of our algorithms, based on modified receiver operating characteristic (mROC) curves.

\subsection{Producing detection and binary maps on test data}\label{sec:testdata}

\begin{figure*}[t]
    \centering
    \includegraphics[width=\textwidth]{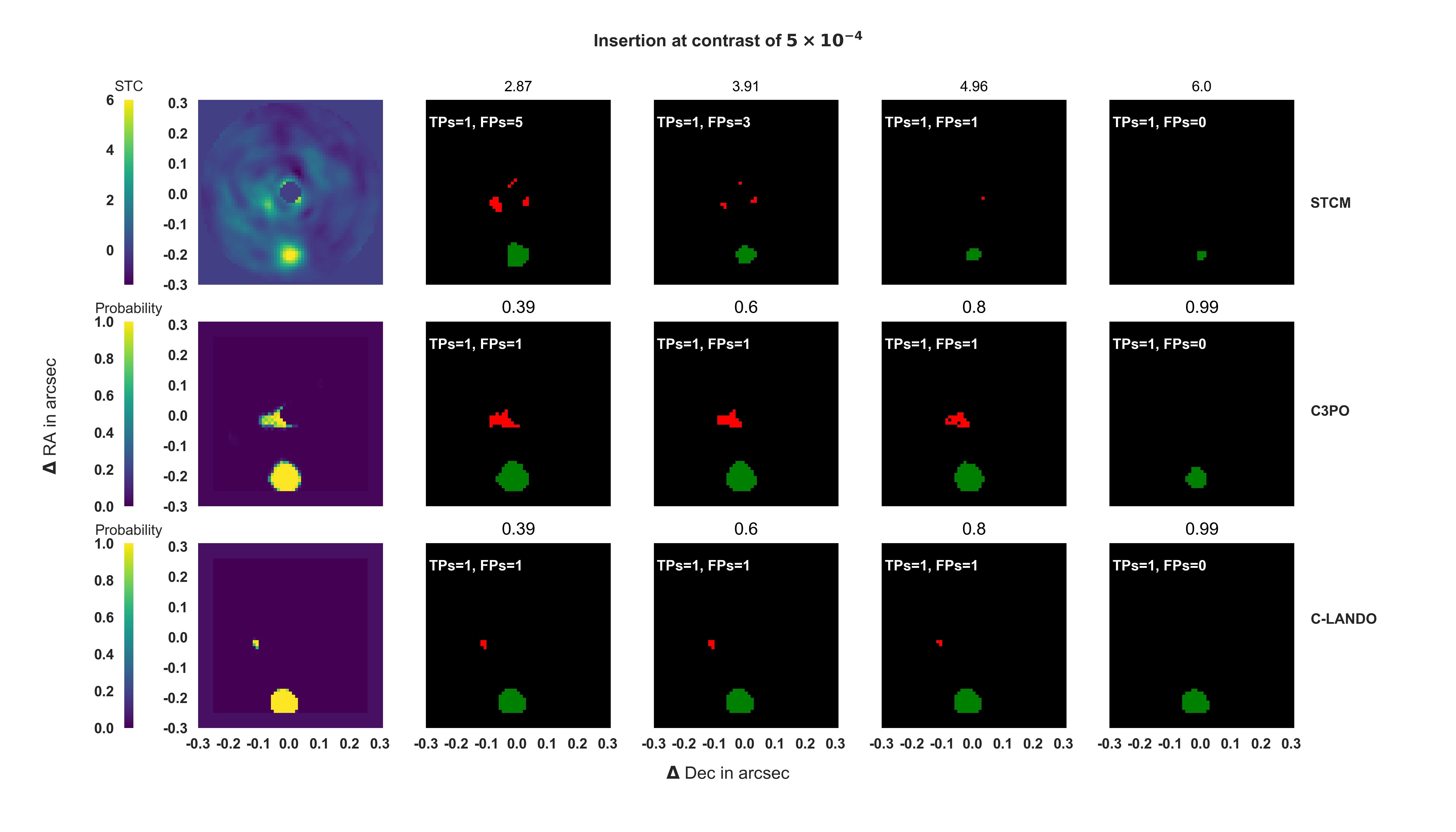}
    \caption{\newchange{Illustration of the TP and FP counting process. Column 1 shows the detection map used to produce four different binary maps, thresholded at different intensity levels (columns 2-5), for the three algorithms (respectively STCM, C3PO, and C-LANDO from top to bottom). A fake companion was inserted at position $\Delta{\rm Dec}=0.0$ and $\Delta{\rm RA} = -0.2$. TPs and FPs are respectively shown in green and in red for the thresholds indicated in the titles of each image. Each red blob or point is counted as a single FP or TP, with blobs representing many connected pixels at the same binary intensity level.
    }}
    \label{fig:sample_detmaps}
\end{figure*}

To produce a test dataset, we choose four PAs corresponding to $60^{\circ},120^{\circ},180^{\circ}$, and $360^{\circ}$.
The choice of these angles is somewhat arbitrary, but it is also done so that all of the tests can be performed with the same angles and there are lesser systematic effects when testing the three algorithms together.
We then go through the procedure of inserting simulated \newchange{young gas giant}s as described in Sect.~\ref{sec: FC insertion}, where we insert \newchange{young gas giant}s at different radial separations over a range of contrasts between $10^{-2}$ to $10^{-4}$. We then produce cross-correlation cubes as described in Sect.~\ref{sec: CC algorithm}, and save these cross-correlation tensors without any further processing. As explained in the previous section, once the algorithms have trained, we pass the test dataset through the neural networks to produce detection maps.
A detection map is a map of the relative weighting of pixels, where noisy pixels carry lower weight and those pixels with an exoplanet carry higher weight.
These maps can be composed using S/Ns, STCMs, or probabilities, and in each case we define thresholds to define which weights constitute a detection (for example $\rm{S/N}\ge 5$).
An example of such detection maps for non-ML algorithms is shown in Fig.~\ref{fig:fig_1}.
Detection maps intrinsically hold the biases of the algorithm that produces them. 
One of the ways of quantifying these biases is the use of  binary maps.
Binary maps are produced when thresholds are applied to a detection map and each pixel above the threshold is assigned a binary value of 1, while the rest of the pixels are set to 0.

We produce detection maps from each of the algorithms as shown in the first column of Fig.~\ref{fig:sample_detmaps}.
Each row in Fig.~\ref{fig:sample_detmaps} starts with the detection map followed by a binary map produced with the threshold indicated in the title of each of the plots in columns two through four.
These detection and binary maps are produced on the test data.
The STCM algorithm (first row in Fig.~\ref{fig:sample_detmaps}) produces a detection map as its natural output.
In order to compare the STCM and ML algorithms, we present the ML algorithms with overlapping patches where the central pixel of each patch is the pixel of interest, whose probability of containing the simulated \newchange{young gas giant} will be predicted by the ML algorithm.
\newchange{This method is thus also adapted to searching for exoplanets in the dataset as we  consider each pixel as the centre of an overlapping patch.}
These probabilities are used to build a detection map.
This constitutes row two and three of the first column of Fig.~\ref{fig:sample_detmaps}.
Columns two to four in Fig.~\ref{fig:sample_detmaps} consist of binary maps produced by thresholding the detection maps.
These binary maps are then used to compute the total number of false positives and true positives for each insertion or each frame (since we have exactly one simulated young gas giant inserted in each frame).

\subsection{Computing true and false positives}
\label{sec:TPFP}

In principle, the threshold is the cutoff that allows us to identify the pixels where the exoplanet is present.
Utilizing thresholding on a detection map facilitates the exploration of the tradeoff between false positives and the detection of faint companions. Lower thresholds, for example, increase the likelihood of identifying faint simulated \newchange{young gas giant}s but concurrently result in a higher incidence of false positives.
Counting the number of false positives and true positives in every binary map and comparing the two values as a function of the threshold allows us to produce classical receiver operating characteristic (ROC) curves.
ROC curves are particularly useful when posing a problem as a two-class classification problem to define the ability of a classifier to make the trade-off between predicting a class and admitting false positives.
They allow to benchmark both ML and non-ML pipelines and make a relative comparison. 

In this paper, we define all the connected pixels in a binary map within $\sim 1$ FWHM of the original insertion of the simulated \newchange{young gas giant} as a true positive (TP) and all other non-zero pixels as a false positive (FP). We perform this analysis over multiple insertions and thereby derive the number of TPs per insertion, also known as a true positive rate (TPR), and the mean number of FPs per detection map, or a mean full frame FP as described in \citet{2018Gomez}. Because we consider the whole field of view to count a mean number of FPs, instead of testing each single resolution element at a time to derive a false positive rate (FPR), we use adapted ROC curves referred to as mROC curves. In comparison to ROC curves, these mROC curves are arguably more relevant to the problem at hand, where the goal is to work at very low FPR (e.g. $3\times 10^{-7}$ for an equivalent Gaussian $5\sigma$ detection). As we produce these mROC curves  curves for each detection map, we mask a circular region of $3.5$ pixels from each frame centre.

\begin{figure*}
\centering
    \includegraphics[width=1.0\textwidth]{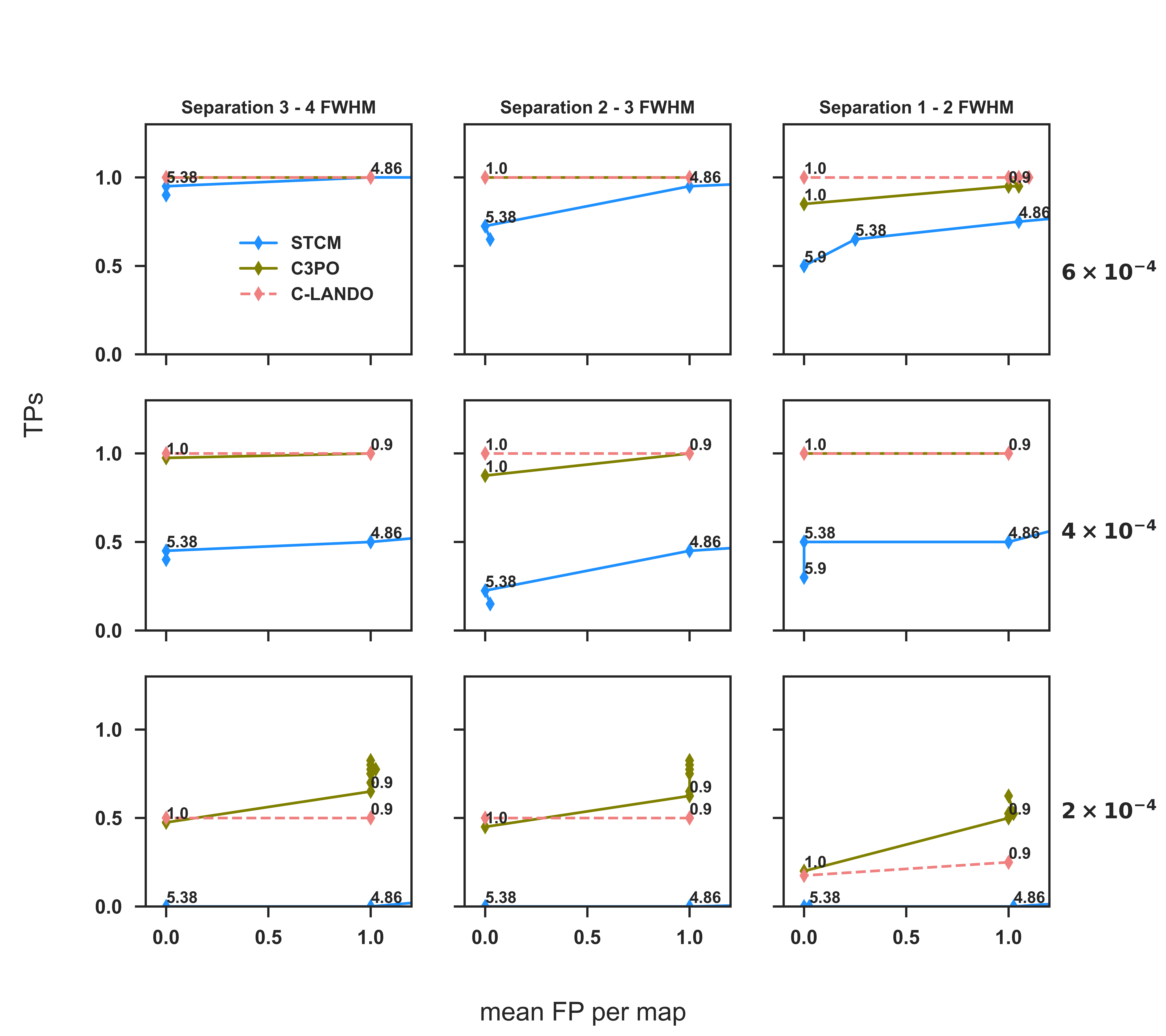}
\caption{ \newchange{mROC curves produced by inserting a total of $40$ exoplanets for a set of contrasts ($6 \times 10^{-4}$, $4 \times 10^{-4}$ and $2 \times 10^{-4}$ from top to bottom) at annuli of different separations in the ranges $3-4$, $2-3$, and $1-2$ FWHM (left, middle, and right columns, respectively).
The x-axis is \textcolor{black}{the} mean FP per map and the y-axis corresponds to \textcolor{black}{the} TPR computed using Eq.~\ref{eq:TPR}. 
The numbers next to each data point correspond to the threshold that was applied to compute the number of TPs and FPs for each binary map (as described in Fig.~\ref{fig:sample_detmaps}).
}}
    \label{fig:fig_4}
\end{figure*}

We use procedures in VIP \citep{2017AJGomezVIP,2023Christiaens} to compute TPs and FPs from a detection map. First, each detection map is subjected to the map-based ROC analysis tool available in VIP. In this tool we set the values of the number of thresholds as $10$, varying from $\approx 1$ to $\approx 6$ for the STCM maps and from $0.1$ to $0.9999$ for the outputs of C3PO and C-LANDO. The FWHM is set to $4.8$ pixels, which is the mean FWHM of the SINFONI instruments in all wavelength bands. 
In order to count the number of TPs and FPs, we use the function \texttt{compute\_binary\_map} described in the VIP documentation\footnote{\url{https://vip.readthedocs.io/en/latest/_modules/vip_hci/metrics/roc.html?highlight=compute_binary_map}}, with the following parameters: the \texttt{overlap\_threshold} variable is set to $0.7$, the \texttt{max\_blob\_fact} variable is set to $3$ and \texttt{npix} variable set to $2$ because of computational errors resulting in slightly different values for higher thresholds even though it is part of the same blob.
We produce such maps for many separations and a range of contrasts.
For every contrast we insert exactly one simulated \newchange{young gas giant} at a specific radial separation and position angle.
Therefore, every detection map will contain at most one TP, and potentially many FPs for the lowest threshold. 
As we progressively increase the threshold the FPs will reduce until at the maximum threshold we have no remaining FP. 
True positives in Fig.~\ref{fig:sample_detmaps} are shown in green and false positives in red.
In order to produce a smooth mROC curve, we repeat this experiment $T_{i}$ number of times, where $T_{i}$ is the product of the total number of radial separations (in this case we pick five separations corresponding to $\approx 1$ FWHM each), the total number of PA that are part of the test set (four angles), and the number of contrast levels at which the injection is performed.
For each insertion, we choose between two contrast values respectively just above and just below the chosen contrast value (e.g. $5\times10^{-4}$ and $3\times10^{-4}$ for a chosen contrast of $4\times10^{-4}$), and inject one companion of each contrast at each position, leading to a total of $40$ insertions to produce one mROC curve.
This results in a total of $40$ detection maps like those in Fig.~\ref{fig:sample_detmaps}, which are then turned into binary maps with various thresholds. 
Each point in Fig.~\ref{fig:fig_4} corresponds to $40$ such binary maps for each algorithm, whose TPR and mean full frame FPs are computed as follows:
\begin{equation}
\rm{TPR} =\dfrac{\sum\limits_{T_i}\rm{TPs}}{T_i} \, , \; 
\rm{mean\; FP} = \dfrac{\sum\limits_{T_i}FPs}{T_i} \, .
\label{eq:TPR}
\end{equation}
It should be noted that the mean number of FPs is always an integer number as the number of FPs remains the same for binary maps of a fixed threshold.
In the unlikely event that a fake companion was inserted at the exact location where a FP would appear for the empty dataset, we still account for it as we compute the mean FP. 
The mean number of FPs is only bounded by the number of pixels in our field-of-view, but we limit the range of the x-axis in our mROC curves to 1, as this is the largest number of FPs produced by the ML algorithm for the considered thresholds, and because this is the regime astronomers are usually interested in.
The numbers around each point in Fig.~\ref{fig:fig_4} represent the threshold for each binary map produced by each algorithm.
We consider the highest threshold as that where ${\rm FP} = 0$.
For the ML algorithms this is satisfied at the highest threshold value of $0.999$, and for the STCM algorithm at a STCM $\sim 6.0$. 

\subsection{Comparing the mROC curves}
\label{sec:roc}

To compare the different ML and non-ML algorithms, we computed their mROC curves for different separations, as each separation corresponds to a different noise regime.
This corresponds to the columns in Fig.~\ref{fig:fig_4}, with the rows corresponding to mean contrasts starting from $6\times10^{-4}$ down to $2\times10^{-4}$.
We have chosen this range of contrasts specifically as it allows us to observe the change in detection sensitivity of the algorithms. The panels start with a similar sensitivity for all algorithms (all the algorithms detect the simulated \newchange{young gas giant} in every insertion with a ${\rm TPR}\approx 1.0$ for the different values of thresholds). 
As we step through different noise levels (left to right) and brightness levels (top to bottom), the STCM algorithm evolves towards a complete insensitivity.
The shape of an mROC curve defines the response of the algorithm to increasing the thresholds and thereby its ability to continue to have TPs while the FPs are removed.
An ideal detection algorithm is where the ${\rm TPR}=1$ even when the threshold changes (i.e. the mROC curve is flat, and lies at the very top of the plot).
This case is illustrated in Fig.~\ref{fig:fig_4}  (cf. first row first column) for all three algorithms and remains the case for the ML algorithms in most of the first two rows, while STCM begins to show a drastic drop in TPR as the simulated \newchange{young gas giant} gets fainter in the second row.
Conversely, when we look at the bottom right set of plots, which correspond to the higher contrast noisy regime, a decrease in threshold will produce more FPs but will also produce higher TPR. 

The performance of the STCM is the baseline that establishes whether a contrast is detectable by non-ML methods. 
This corresponds to contrasts in the top left of the plot, particularly the top left extreme where all three algorithms perform similarly in terms of the TPs \textcolor{black}{i.e., ${\rm TPR}=1$ for ${\rm FP}=0$}.
Thus, for this paper, we study those contrasts at which STCM produces ${\rm TPR}<1$ and compare how ML algorithms perform for these contrasts and separations.
As we go from the top left towards the bottom right we see TPRs going down across all algorithms, until we reach a TPR $=0$ at the bottom row for STCM. 
In general, ML algorithms continue to detect the majority of simulated \newchange{young gas giant}s (TPR $\ge 0.5$) other than for the highest contrast companion closest to the frame centre.
For the lowest contrast, we see that STCM produces a TPR $=0$ for almost all separations and values of the thresholds.
Conversely, the ML algorithms continue to be able to produce a TPR $>0$ for all the considered contrasts.
The additional advantage offered by ML algorithms is that the FPs do not seem to increase indefinitely, whereas with STCM it is obvious that the FPs will continue to increase at lower thresholds.
An interesting observation is that the brightness of the \newchange{young gas giant} seems to be the single most important astronomical aspect which impacts detectability with ML algorithms.
We see that the TPR does not drop when reducing radial separation (i.e. across columns in Fig.~\ref{fig:fig_4}), but rather drops with increasing contrast. 
Conversely,  the brightness of the companion and the background stellar contamination seems to both be impacting factors for the STCM, as we observe a drop in TPR both with increasing contrast and reducing radial separations (i.e. both rows and columns in Fig.~\ref{fig:fig_4}).


\section{Discussion}
\label{sec:discussion}

The results in the previous section indicate that ML algorithms provide an advantage in noisier regimes, closer to the frame centre and particularly with higher contrast exoplanets.
But the reality is that the algorithm uses data features (temporal, spatial and velocity) diversity naturally present in the data efficiently and consistently in the test data.
One of the questions that is raised is which features or dimensions of the data are producing this high detection sensitivity and robustness that make them more reliable?
We also have several dimensions such as the \textcolor{black}{spatial}, the cross-correlation velocity dimension, and the temporal dimension.
In this section, we analyse this question in the light of our data and results.
We start by discussing the reasons for the relatively higher TPR produced by C3PO when compared to C-LANDO.
This will serve to illuminate as well, why the temporal dimension does not lend much more detection sensitivity than just using the image dimension alone.
We  then make a test of the same data both with and without the velocity to provide evidence of the importance of the velocity dimension to our results.

\subsection{Impact of the temporal dimension}

\begin{figure*}
\centering
\includegraphics[width=\textwidth]{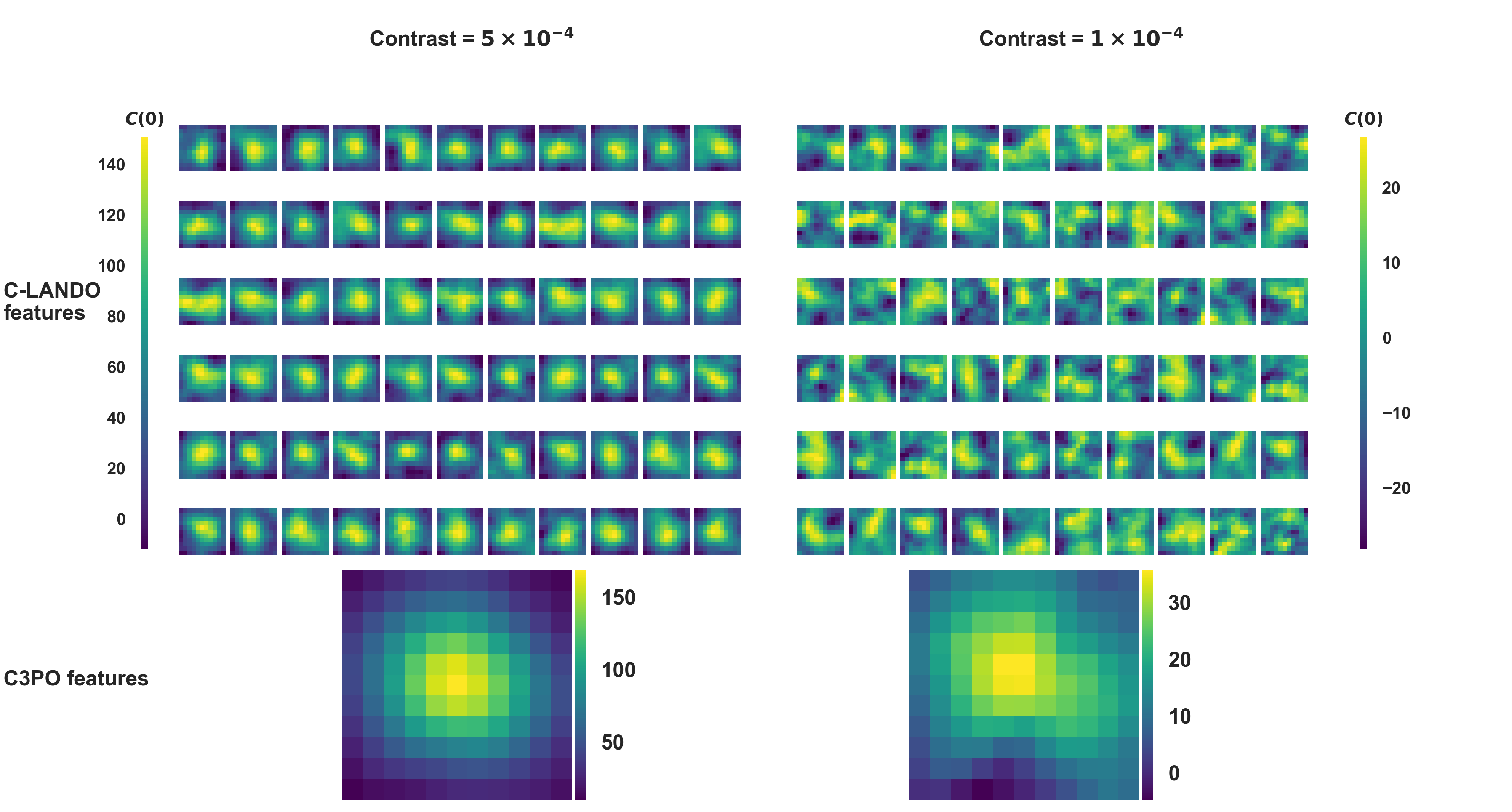}
\caption{Data presented to the ML algorithms for training, for two different contrast levels. The upper rows of miniatures depict the spatial distribution of the cross-correlation signature along the temporal dimension, used for C-LANDO training, while the lower images depict the temporal mean of the same cube, used for C3PO training. }
\label{fig:disc-part1}
\end{figure*}

The previous section showed us that C3PO is more sensitive both at higher contrast and noisier regimes than C-LANDO.
We posit that one of the reasons for this higher detection sensitivity is the consistency in the presence of signal in the input cubes to C3PO.
While both C3PO and C-LANDO have been trained with the same data and are considered equally well trained, the difference in performance is a measure of the features necessary to make a prediction being present in the data.
Therefore, it appears that while for lower contrasts \textcolor{black}{(i.e., brighter exoplanets)} C-LANDO sees the signature of the exoplanet being consistently present, in the temporal dimension this behaviour changes with higher contrast insertions.
We use Fig.~\ref{fig:disc-part1} to explain this behaviour.
This figure shows the input to both algorithms, composed of cross-correlation tensors that are cropped to the 2-FWHM patch tensors.
In the first six rows of the figure, we depict patches of some of the $83$ temporal cubes (i.e. the temporal features that are used to train C-LANDO), while the final row shows the mean of these temporal cubes (i.e. the features used to train C3PO).
The two large columns represent two different contrasts at which the simulated \newchange{young gas giant} is inserted. On the left is a contrast corresponding to the first column second row of Fig.~\ref{fig:fig_4}, where both algorithms perform very similarly.
Both in amplitude and spatial features the first large column shows quite similar features, where C-LANDO just seems to have an additional $83$ copies of the features that C3PO receives. 
The data features the algorithm looks for is the spatial \textcolor{black}{distribution} of the cross-correlation coefficients, as much as \textcolor{black}{signal consistency} in the temporal dimension.
The second large column corresponds to a \newchange{young gas giant} inserted for the same separation at a higher contrast.
Thus, the only difference between the columns is the brightness of the simulated \newchange{young gas giant}.
In this column the first difference that we notice is the difference in amplitude between the first six rows and the last row.
The first six rows have an amplitude range closer to the class $C_{\rm N}$ in Fig.~\ref{fig:fig-2}, whereas the final row seems to have linearly scaled amplitude to its lower contrast \newchange{(brighter)} counterpart (first column).

The spatial signature of the cross-correlation plays another, in our view, crucial role in the higher TPR produced by C3PO for higher contrasts.
When we compare the inputs to both the algorithms between the different contrasts, the spatial distribution of the cross-correlation coefficients is better conserved for the input to C3PO than for C-LANDO, although the mean cross-correlation strength is reduced for a higher contrast.
This seems to be another key differentiating factor between the two algorithms where it appears that computing the mean to produce the last row focuses the features better to enable an ML algorithm to be more sensitive to a high-contrast exoplanet.
\subsection{Importance of the velocity dimension}
\begin{figure*}
    \centering
    \includegraphics[width=\textwidth]{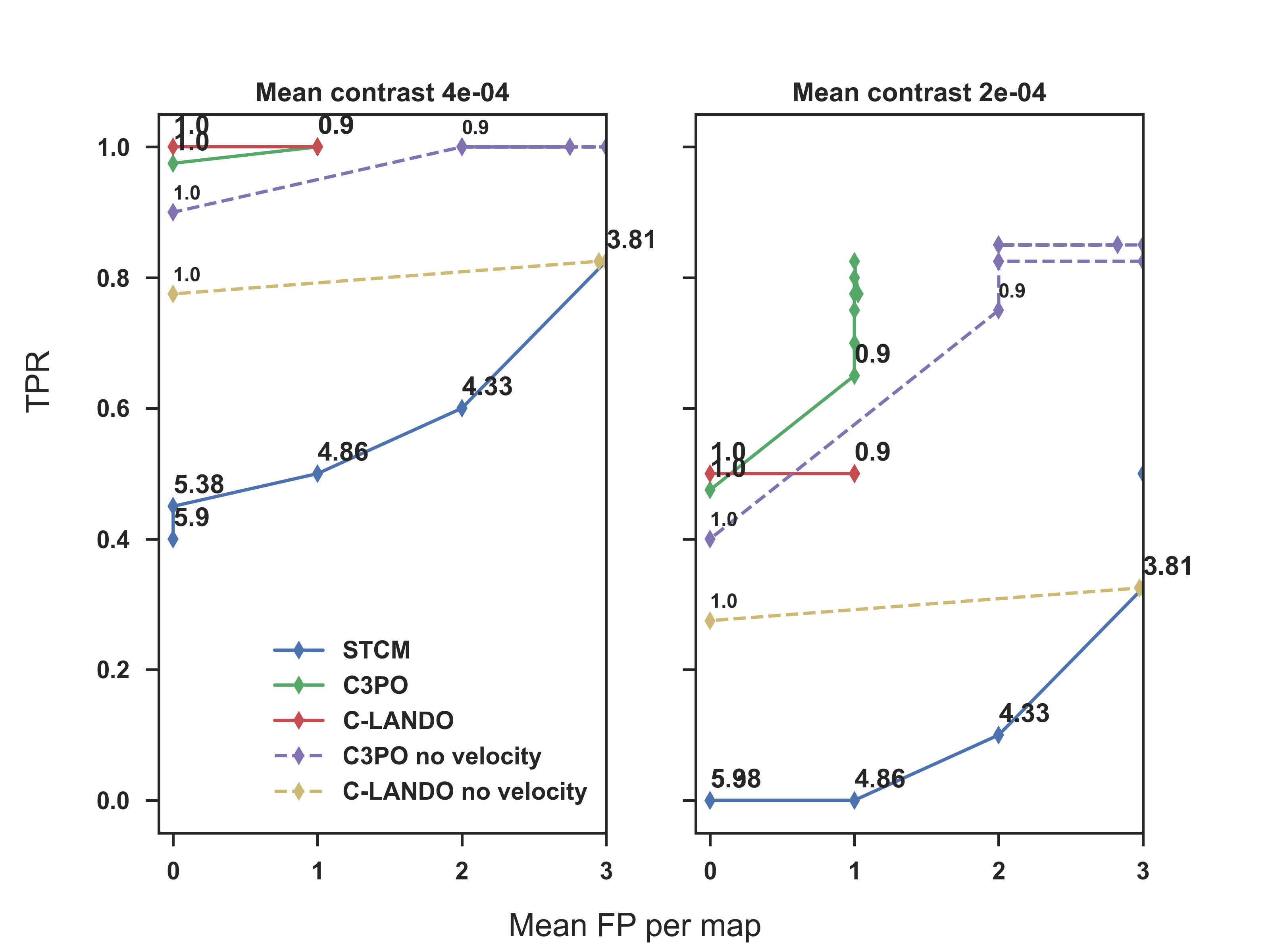}
    \caption{\newchange{Comparison of mROC curves produced with and without using the velocity dimension, using otherwise the same dataset, at two contrast levels ($4\times 10^{-4}$ and $2\times 10^{-4}$) and for a separation of 3 to 4 FWHM. In the  cases where the velocity dimension is not used, illustrated with the purple and yellow dashed lines for C3PO and C-LANDO, respectively, the velocity dimension is replaced by the central velocity bin. This makes the comparison to STCM a little more balanced as STCM does not have the capability to use the velocity dimension.
A final noteworthy point is that given the data dimensionality used by both the ML algorithms and STCM, it is truly fair to compare the solid blue line in Fig.~\ref{fig:novel_roc} and the dashed lines produced by the ML algorithms, and we note the TPRs produced by the ML algorithms are higher; however, the FPs produced by C-LANDO are similar, when reducing the threshold, to those produced by STCM (${\rm FP}=3$). 
    We note that the solid curves remain the same as in Fig.~\ref{fig:fig_4}.}
    }
    \label{fig:novel_roc}
\end{figure*}

The importance of how spatial features are driving the higher performance of C3PO was illustrated in the \textcolor{black}{previous} section. 
The STCM algorithm sees the same features as C3PO, but \textcolor{black}{with $v=0$, and thus only sees the $C(0)$ coefficient}.
Since velocity seems to be the key feature difference between C3PO and STCM, it is relevant to discuss the importance of the velocity feature in the data. 
In order to appreciate the importance of this dimension, we train the ML algorithms both with the full range of the velocity shift (i.e. between $-100$ and $100$ km/s) to produce ROC curves in Fig.~\ref{fig:fig_4}, but also train them by choosing only the velocity where the cross-correlation value is the maximum (i.e. $v=5$ km/s in Fig.~\ref{fig:fig-2})
We note that because of the medium resolution of SINFONI, there is no significant difference between the spatial distribution and mean value of cross-correlation between $v=-5$, 0, and 5 km/s. 
Then, we re-create C3PO and C-LANDO such that the value for the number channels is $1$ instead of $20$ as in Fig.~\ref{fig:c3po schematic}.
All other dilations/changes are appropriately altered automatically as a consequence of this but the hyperparameters (i.e. number of neurons, layers, the mix of contrasts used for training, optimization algorithm, and error function) remain the same.
We perform training exactly in the manner described in Sect.~\ref{sec:ML algorithms}, compute the TPs and FPs in frame exactly as in Sect.~\ref{sec:results}, and then produce sample ROC curves in Fig.~\ref{fig:novel_roc}, where the two plots are obtained for insertions of simulated \newchange{young gas giant}s at different contrasts but at a fixed separation from the frame centre.
The dashed lines in Fig.~\ref{fig:novel_roc} are those produced with no explicit velocity evolution in the data.
The solid lines are the same as those produced in Fig.~\ref{fig:fig_4}.

The first noteworthy point is that we have extended the FP axis up to $3$ FPs, because both ML algorithms without access to the velocity dimension produce more FPs.
The second point is that the evolution (i.e. shape of the mROC curves) of the TPRs produced by the ML algorithms remains similar both with and without the velocity dimension between two different contrasts (i.e. the TPR falls by a fraction of $\sim 50\%$ between contrasts).
However, within the same contrast regime (in this case each sub-plot), the TPR produced by C-LANDO with velocity is $\sim 20\%$ higher than without the velocity evolution (for a fixed FP), whereas the evolution of the TPR produced by C3PO is less dramatic.
The velocity dimension produces a distinct systematic variation of spatial features (i.e. both cross-correlation values and the spatial distribution of the cross-correlation).
This is becomes evident in Fig.~\ref{fig:fig-2}, when comparing the evolution of the spatial features with velocity between $C_{\rm FC}$ and $C_{\rm N}$.
This produces an additional dimension that can be relied on to make this distinction between different pixels, making the detection less confused ergo more sensitive.
This should particularly be useful when one of the dimensions used are not able to make these distinctions for a specific insertion.
Therefore, while introducing the velocity dimension, C-LANDO produces an improvement in both TPR and FPs, the improvement in C3PO is mostly confined to having fewer FPs.

A final noteworthy point is that given the data dimensionality used by both the ML algorithms and STCM, it is truly fair to compare the solid blue line in Fig.~\ref{fig:novel_roc} and the dashed lines produced by the ML algorithms, and we notice that the TPRs produced by the ML algorithms are higher, however the FPs produced by C-LANDO are similar, when reducing the threshold, to those produced by STCM (${\rm FP}=3$). 
When we consider the performance of the ML and non-ML algorithms, it is clear that even with the same data dimensions C3PO outperforms STCM in detection sensitivity.
\section{Conclusion}

In this paper we explored the detectability of \newchange{young gas giant}s in high-contrast imaging data, while having access to medium-resolution spectra for each pixel of the image, with several such cubes produced temporally.
The fundamental question we   addressed is whether the use of such data with ML algorithms produces any improvement in detection sensitivity to \newchange{young gas giant}s.
In order to efficiently use the spectral dimension, we   used cross-correlations to produce cross-correlation tensors, which were then subjected to further statistical analysis to determine if they contain an injected exoplanet.
We used ML and non-ML algorithms that are capable of detecting the cross-correlation signature of an exoplanet.
In order to answer this question, we  used ROC curves as a method to identify the trade-off between incorrectly labelling speckles as exoplanets in test data and detecting high-contrast exoplanets.
To build ROC curves, we inserted simulated \newchange{young gas giant}s in a dataset devoid of any known companions, at different radial separations from the host star to simulate different noise regimes and at different contrast to simulate exoplanets of different brightness.
To detect these simulated \newchange{young gas giant}s we  devised a non-ML algorithm (STCM), a CNN-based algorithm (C3PO), and a CNN-LSTM-based algorithm (C-LANDO). 
We trained the ML algorithms on part of the data and reserved the rest to test and benchmark our algorithms. 
We trained our algorithms so that the validation set had the  fewest  FPs, and tuned our algorithms accordingly. 

The most notable results of our study are the following:
\begin{itemize}
    \item The ability to detect TPs reduces for the STCM algorithm for companions closer to the frame centre, while it remains relatively unchanged for ML algorithms.
    \item The ability of ML algorithms to harness additional dimensionalities such as velocity seems to be at the root of their increased sensitivity.
    \item The use of the temporal dimension does not seem to provide an added value in terms of detectability.
\end{itemize}

There are a few limitations to this work that merit a mention.
The first  is related to the use of a unique type of exoplanet, while the use of cross-correlation also allows the use of multiple templates for multiple types of exoplanets.
A companion paper \citet{2024Garvin} addresses this topic, where they show that it is possible to account for template variability by incorporating several spectral templates as convolution filter depth for CNNs. 
This   accounts for uncertainties in the planet's atmospheric characteristics when performing detection in spectral data.
To increase this flexibility, they also use individual molecular line lists instead of full atmospheric templates.
Another limitation of this work is the limited data that was used for this study, which means that the selection of test data reduced the amount of data available to train the algorithms.
Despite this significant limitation, the ML algorithms still continued to be more sensitive to high-contrast exoplanets, even at smaller radial separations without any evidence of overfitting.
If our algorithms overfit the training data, we typically expect to see a large number of FPs in the test data, but very few in the training data. 
The fact that this has performed well on this dataset, which is very noisy and lacks a coronagraph to enable high-contrast imaging, gives us confidence in a future iteration to try this algorithm on a less noisy dataset.
Finally, ML algorithms have evolved rapidly in the last few years, with the advent of Vision Transformers \citep{2020ViT} and recipes to train them on smaller datasets \citep[e.g.][]{2022Gani}. Those algorithms have recently been shown to be very effective on the same data dimensions, and would be a direction to explore for future research. Although we used relatively standard network architectures, this paper sets the methodology to leverage a computer vision algorithm effectively with spectral and temporal dimensions.

\begin{acknowledgements}
R.N.R., O.A., and V.C. are funded by the Fund for Scientific Research (F.R.S.-FNRS) of Belgium. 
R.N.R was funded for this project through the FRIA grant N\textsuperscript{o}   40004011. This project has received funding from the European Research Council (ERC) under the European Union's Horizon 2020 research and innovation programme (grant agreement No 819155). R.N.R also wishes to thank the members of PSILab headed by O.A and Dr Sascha Quanz for the kind inputs and discussions. E.O.G gratefully acknowledges the financial support from the Swiss National Science Foundation (SNSF) under project grant number 200020\_200399.
\end{acknowledgements}
\bibliographystyle{aa}
\bibliography{references_oab.bib}

\end{document}